\documentclass{ws-ijmpd}
\usepackage{graphicx}
\usepackage{bm}
\usepackage{subfigure}
\usepackage{longtable}
\usepackage{amssymb}
\usepackage{amsmath}
\begin{document}
\markboth{D. M\"uller and J. A. de Deus}{Bianchi $I$ solutions of effective gravity}
\title{Bianchi $I$ solutions of effective quadratic gravity}

\author{Daniel M\"uller\footnote{Email: muller@fis.unb.br} }
\address{Instituto de F\'\i sica, UnB Campus Universit\'ario Darcy Ribeiro Cxp 04455\\ 70919-970, Bras\'\i lia DF 
Brazil  \\ }
\author{Juliano A. de Deus \footnote{Email: julianoalves@fis.unb.br}}
\address{Instituto de F\'\i sica, UnB Campus Universit\'ario Darcy Ribeiro Cxp 04455 \\ 70919-970, Bras\'\i lia DF 
Brazil}
\maketitle

\begin{history}
\received{Day Month Year}
\revised{Day Month Year}
\comby{Managing Editor}
\end{history}

\begin{abstract}
It is believed that soon after the Planck time, Einstein's general relativity theory should be corrected to an effective quadratic theory. Numerical solutions for the anisotropic generalization of the Friedmann ``flat" model $E^ 3$ for this effective gravity are given. It must be emphasized that although numeric, these solutions are exact in the sense that they depend only on the precision of the machine. 
The solutions are identified asymptotically in a certain sense. It is found solutions which asymptote de Sitter space, Minkowski space and a singularity. This work is a generalization for non diagonal spatial metrics of a previous result obtained by one of us and a collaborator for Bianchi $I$ spaces. 
\end{abstract}

\section{Introduction}
\newcommand{\imsize}{1.0\columnwidth} 
\newcommand{\halfsize}{0.50\columnwidth} 
The semi-classical theory considers the back reaction of quantum fields in a classical geometric 
background. It began about forty years ago with de Witt\cite{DeWitt}, and since then, its 
consequences and applications are still under research, see for example\cite{Hu}.

Different from the usual Einstein-Hilbert action, the one loop effective gravitational action surmounts 
to quadratic theories in curvature, see for example Refs. \refcite{DeWitt}, \refcite{grib,bd}. It is the gravitational version of the Heisenberg-Euler electromagnetism. As it is well known, vacuum polarization introduces non linear corrections into Maxwell electrodynamics, see Schwinger\cite{schwinger}, first obtained by Heisenberg and Euler\cite{heisenberg}.

This quadratic gravity was studied in the pioneering works of \cite{buch,ruz} and \cite{S,lukash,zs}. Starobinsky idea was that the higher order terms describe a slowly decaying dark energy, i.e. to produce slow-roll inflation. It is of interest for example in the context of the final stages of evaporation of black holes,  inflationary theories\cite{coule}, in the approach to the singularity\cite{montani} and also in a more theoretical context\cite{cotsakis}.

The effective gravity was apparently first
investigated by Tomita \cite{tomita} for general Bianchi $I$
spaces. They found that the presence of anisotropy contributes to the 
formation of the singularity. In Ref. \refcite{gurovich} it is shown that the equations for the anisotropic part of the Bianchi type I metric can be integrated for a general $f(R)$ theory including the quadratic $R+R^2$ gravity. 
Ref. \refcite{berkin} shows that a quadratic
Weyl theory is less stable than a quadratic Riemann scalar $R^2$. Homogenous solutions in the context of quadratic gravity was also addressed by\cite{b1,b2,bh,clifton}. Schmidt\cite{hjs} does a very interesting  
review of higher order gravity theories in connection to cosmology. 

We have previously numerically investigated Bianchi $I$ type solutions\cite{sandro} for the diagonal line element only. Soon after this 
result we also investigated the stability of the this particular Bianchi $I$ case\cite{daniel-sandro}, also for the diagonal line element. We must mention that the question of stability of the de Sitter solution in quadratic gravity was previously addressed by \cite{muller}. It 
turns out that for zero cosmological constant $\Lambda=0$, Minkowski geometry is structurally stable in 
the sense that there is a basin of attraction to Minkowski solution.  And for $\Lambda >0 $ de Sitter type 
geometry is structurally stable, also in the same sense. Thus, soon after the Planck era, the effective 
one loop quadratic gravity predicts for the very particular Bianchi I models, that there is a basin of 
attraction to Minkowski space for $\Lambda=0$ and a basin of attraction to de Sitter space for $\Lambda>0$, see\cite{daniel-sandro}.  

The purpose of this work is to consider more general geometries than the ones considered in\cite{sandro} and\cite{daniel-sandro}, for instance with a non diagonal line element.  Bianchi $I$ is the anisotropic generalization 
of the Euclidean $E^3$ Friedmann ``flat" space. The time like vector is orthogonal to the group orbit, geodesic and with zero vorticity.
For an excelent and recent review in Bianchi cosmologies, see Ref.  \refcite{BiancRevisEllis}.

For any spatially homogenous space-time, the 
dynamical equations of motion, of any metric theory, result in a non linear system of ordinary differential equations. 
For the particular quadratic gravity, the ordinary differential equation is of degree $4$. 

The initial conditions chosen are very near to exact solutions of the quadratic theory, and also only classical vacuum source is considered, which we believe is the best description soon after the Planck era. The solutions are obtained numerically and they are understood asymptotically in a certain sense described in the text. It is shown that depending on the initial conditions and parameters, the de Sitter solution, or Minkowski solution, or a singularity type solution is obtained asymptotically. 

The paper is organized as follows. Section $2$ presents the field equations for the Bianchi $I$ case, and also the Newman-Penrose coefficients. In Sec $3$, three different initial conditions are analyzed: one that converges to de Sitter space, one that converges to Minkowski space, and one that converges to a singularity. We believe the index notation should be clear from the text. 

\section{Theory and Development}

The field equations for the semiclassical theory, are obtained performing
metric variations in the gravitational Lagrangian 
\begin{equation}
\mathcal{L}=\sqrt{-g}\left[-\Lambda+R+\alpha\left(R_{ab}R^{ab}-\frac{1}{3}R^{2}\right)+\beta R^{2}\right]+\mathcal{L}_{q},
\label{acao}
\end{equation}
where $\mathcal{L}_{q}$ is the quantum part of Lagrangian and $\alpha$
and $\beta$ are constants. For the spatially homogenous space they are described
by the tensor $E=E_{ab}\omega^{a}\otimes\omega^{b}$, $\omega^4=dt$,
\begin{equation}
E_{ab}\equiv G_{ab}+\frac{1}{2}g_{ab}\Lambda-\left(\beta-\frac{1}{3}\alpha\right)H_{\: ab}^{(1)}-\alpha H_{\: ab}^{(2)}=0,\label{eq field equation for semiclassical theory}
\end{equation}
where
\begin{eqnarray*}
&&G_{ab}=R_{ab}-\frac{1}{2}g_{ab}R,\\
&&H_{ab}^{(1)}=\frac{1}{2}g_{ab}R^{2}-2RR_{ab}-2g_{ab}\nabla_c\nabla^c R+2R_{;ab},\\
&&H_{ab}^{(2)}=\frac{1}{2}g_{ab}R^{cd}R_{cd}-\nabla_c\nabla^c R_{ab}-\frac{1}{2}g_{ab}\nabla_c\nabla^c R\\
&&+R_{;ab}-2R^{cd}R_{cbda}.
\end{eqnarray*}

The counterterms $R^2$, $R_{ab}R^{ab}$, $\Lambda$  and $R$ in (\ref{acao}) are precisely the ones necessary in order to obtain a finite vacuum expectation  value of the energy momentum tensor, see for example Ref. \refcite{christensen}. A theory without these counterterms is inconsistent from the point of view of the renormalization of the quantum field in $\mathcal{L}_{q}$. The renormalized vacuum expectation value of the energy momentum tensor is set to zero, which emphasizes the effects of a theory that should have been considered from the start. We are disregarding any classical contribution in this work. 

The finite contributions to the vacuum expectation value of the energy momentum tensor are known only for very particular situations, for example, split rank spaces, for which the heat kernel can be obtained exactly \cite{camporesi}, then the point splitting method gives $\langle 0|T_{ab}| 0 \rangle$ exactly. These solutions are known as self consistent, $G_{ab}=\langle 0|T_{ab}| 0 \rangle$ and there are very few cases, for example Ref. \refcite{dowker}. For another very interesting example of a self consistent anisotropic solution supported by vacuum polarization see  Ref. \refcite{kofman}.

As in any other metric theory, the covariant divergence of $E_{ab}$
must be zero
\[
\nabla^{a}E_{ab}=0.
\]
From $\nabla^{a}E_{ab}=0$ it follows that if $E_{44}=0$ and $E_{4\alpha}=0$
initially, then they will remain zero for any time, and act as constraints
on the initial conditions. Consequently, these constraints are checked
to test the accuracy of the numerical results, while $E_{\alpha\beta}=0$
represent the real dynamical equations of the problem (see for example
Ref. \refcite{Stephani}, p. 165).

The homogeneous and anisotropic metric is supposed in the following
way,\begin{equation}
g_{ab}(t)=\left(\begin{array}{cccc}
a_{1}^{\:2}(t)&0 & 0 & 0\\
0 & a_{2}^{\:2}(t) & a_{4}(t)&0\\
0 & a_{4}(t) & a_{3}^{\:2}(t)&0\\
0&0 &0&-1
\end{array}\right),\label{eq homogeneous and anisotropic line element}\end{equation}
\begin{equation}
ds^{2}=-dt^{2}+g_{\alpha\beta}(t)dx^{\alpha}dx^{\beta},\label{eq line element for spatially homogeneous four-space again}\end{equation}
where $g_{\alpha\beta}$ is the spatial part. The replacement
of the above line element in (\ref{eq field equation for semiclassical theory})
results in a non-linear fourth-order ordinary differential equation
system in the functions $a_{i}(t)$, $i=1,2,3,4$:
\begin{equation}
\frac{d^{4}}{dt^{4}}a_{i}(t)=f_{i}\left(\frac{d^{3}}{dt^{3}}a_{j}(t),\ddot{a}_{j}(t),\dot{a}_{j}(t),a_{j}(t)\right).
\end{equation}

First let us emphasize that every Einstein space, satisfying $R_{ab}=\Lambda g_{ab}/2$, is an exact solution of this effective theory given in (\ref{eq field equation for semiclassical theory}). Note the particular case when the constant $\Lambda=0$: Vacuum solutions of Einstein's equations are also exact solutions of (\ref{eq field equation for semiclassical theory}). So there is the following exact solution of (\ref{eq field equation for semiclassical theory})
for Bianchi $I$ with the diagonal metric $g_{\alpha\beta}(t)=\mbox{diag}(a_{1}^{\:2}(t),a_{2}^{\:2}(t),a_{3}^{\:2}(t))$
\begin{eqnarray}
&&a_1(t)=Ce^{t\sqrt{\Lambda/6}}\nonumber\\
&&a_2(t)=Ce^{t\sqrt{\Lambda/6}}\nonumber\\
&&a_3(t)=Ce^{t\sqrt{\Lambda/6}}, \label{soldeSitter}
\end{eqnarray}
where $C$ is an integration constant. The particular case when $\Lambda\rightarrow 0$, is of course Minkowski space. 

The intention is to characterize the obtained numerical solutions in some way. The Weyl tensor follows from the Riemann tensor as
\begin{eqnarray}
&&C_{abcd}=R_{abcd} -\frac{1}{2}(g_{ca}R_{bd}+g_{db}R_{ca}-g_{cb}R_{da}-g_{da}R_{cb})\nonumber\\
&&+\frac{1}{6}R(g_{ca}g_{db}-g_{cb}g_{da}).
\label{weyl-riemann}
\end{eqnarray}

A complex null basis can be defined 
\begin{eqnarray}
&&k^{a}k_{a}=k^{a}t_{a}=k^{a}\bar{t}_{a}=l^{a}l_{a}=l^{a}t_{a}=l^{a}\bar{t}_{a}=t^{a}t_{a}=\bar{t}^{a}\bar{t}_{a}=0,\nonumber\\
&&t^{a}\bar{t}_{a}=-k^{a}l_{a}=1\label{basenula},
\end{eqnarray}
with the corresponding null metric 
\begin{equation}
\tilde{g}_{AB}=g_{ab}A^aB^b=\left(\begin{array}{cccc}
0&-1 & 0 & 0\\
-1& 0 & 0&0\\
0 & 0&0&1\\
0&0 &1&0
\end{array}\right),
\label{metricanula}
\end{equation}
where $A^a$ and $B^b$ are the null vectors in (\ref{basenula}). The Newman-Penrose complex coefficients are in fact the tetrad components of the 
Weyl tensor 
\begin{eqnarray*}
&&\psi_{0}=C_{abcd}k^{a}t^{b}k^{c}t^{d},\\
&&\psi_{1}=C_{abcd}k^{a}l^{b}k^{c}t^{d},\\
&&\psi_{2}=C_{abcd}k^{a}t^{b}\bar{t}^{c}l^{d},\\
&&\psi_{3}=C_{abcd}k^{a}l^{b}\bar{t}^{c}l^{d},\\
&&\psi_{4}=C_{abcd}\bar{t}^{a}l^{b}\bar{t}^{c}l^{d}.
\end{eqnarray*}
Also the Ricci tensor can be projected to the null tetrad, resulting into the independent components,
\begin{eqnarray}
&&R_{kk}=R_{ab}k^ak^b\nonumber\\
&&R_{kl}=R_{ab}k^al^b\nonumber\\
&&R_{ll}=R_{ab}l^al^b\nonumber\\
&&R_{kt}=R_{ab}k^at^b\nonumber\\
&&R_{lt}=R_{ab}l^at^b\nonumber\\
&&R_{tt}=R_{ab}t^at^b\nonumber\\
&&R_{t\bar{t}}=R_{ab}t^a\bar{t}^b, \label{riccinull}
\end{eqnarray}
the first $3$ and the last one are real, and the remaining ones are arbitrary complex numbers comprising of course 10 independent components. 

The Newman-Penrose coefficients are related to the Petrov classification as shown for example
in Refs. \refcite{ExactSolutionsStephani} and \refcite{Stephani}. When all the $\psi$'s
are zero it's a Petrov type O, which characterizes a conformal Minkowski  
space and the Weyl tensor vanishes. An exactly conformally flat solution is either a generalized Friedmann solution or an interior Schwarzschild solution as it can be seen for example in Ref. \refcite{ExactSolutionsStephani} pg. 413. Minkowski  and de Sitter geometries, are particular cases of conformally flat solutions. 

Before getting into the numeric result, we call the attention to the book\cite{dsc}, pp. 62- 64. As it is well written there, there is not a complete statement as to what constitutes a minimal set for ensuring that a cosmological model is close to Friedmann-Lemaitre model. This book is based on an article by Stoeger {\it et al}. \cite{Stoeger1995} for which some assumptions are made: i) Einstein's equations are satisfied. ii) the source is a mixture of radiation and dust. 

We are not concerned if the two above conditions are satisfied in this present work. So we do not expect that their result, albeit being very interesting, should be verified in the particular context being discussed here. 

In the following we will present a particular example. Consider the Bianchi $I$ metric written in the appropriate coordinate base $dx,\, dy,\, dz,\, dt$
\begin{equation}
g_{ab}=\left(\begin{array}{cccc}
a(t)^2&0 & 0 & 0\\
0& b(t)^2 & 0&0\\
0 & 0&b(t)^2 &0\\
0&0 &0&-1
\end{array}\right).  
\label{cexemplo}
\end{equation}
The time like vector $u^a=(0,0,0,1)$ is geodesic and orthogonal to the group orbit. The magnetic part $H_{ab}=0$, and the electric $E_{ab}$ part of the Weyl tensor is
\[ 
E_a^b=\left(\begin{array}{cccc}
2\psi_2&0 & 0 & 0\\
0& -\psi_2 & 0&0\\
0 & 0&-\psi_2 &0\\
0&0 &0&0
\end{array}\right),
\]
where $\psi_2$ is the above defined Newman-Penrose coefficient 
\begin{equation}
\psi_2=\frac{1}{6}\left( \frac{-(\dot{b}(t))^2+b(t)\ddot{b}(t)}{b(t)^2}\right).
\label{psi2}
\end{equation}
Also regarding (\ref{cexemplo}), the expansion $H=1/3\Theta$, $\Theta=\nabla_c u^c$ is
\begin{equation}
H=\frac{1}{3}\frac{\dot{a}(t)}{a(t)}+\frac{2}{3}\frac{\dot{b}(t)}{b(t)}.
\label{expansao}
\end{equation}
Now consider 
\begin{eqnarray}
a(t)=\frac{A}{1-t^{-1/3}\sin (t)} \nonumber\\
b(t)=\frac{A}{1-t^{-1/4}\cos (t)}.
\label{fcexemplo}
\end{eqnarray}
From (\ref{fcexemplo}) it is easily seen that (\ref{cexemplo}) as $t\rightarrow \infty$ is
\begin{equation}
\lim_{t\rightarrow \infty}g_{ab}=\left(\begin{array}{cccc}
A^2&0 & 0 & 0\\
0& A^2 & 0&0\\
0 & 0&A^2 &0\\
0&0 &0&-1
\end{array}\right),
\label{cexassint}
\end{equation}
which is Minkowski geometry. 

Considering (\ref{fcexemplo}) and (\ref{cexemplo}), now we can explicitly obtain the asymptotic $t\rightarrow \infty$ limits of the Newman-Penrose coefficient (\ref{psi2}) and the expansion (\ref{expansao}) to be 
\begin{eqnarray}
&&\lim_{t\rightarrow \infty}\psi_2=-\frac{1}{6}t^{-1/4}\sin (t)\nonumber\\
&&\lim_{t\rightarrow \infty}E_{ab}E^{ab}=\frac{1}{6}t^{-1/2}\sin (t)^2\nonumber\\
&&\lim_{t\rightarrow \infty} H=\frac{2}{3}t^{-1/4}\cos (t)
\label{psi,H,assintoticos}
\end{eqnarray}
 
This example shows that  Definition 2.1 of \cite{dsc} 
\begin{eqnarray}
\sqrt{H_{ab}H^{ab}}/H^2<<\epsilon\nonumber\\
\sqrt{E_{ab}E^{ab}}/H^2 << \epsilon 
\label{cellis}
\end{eqnarray}
$0<\epsilon<<1$ is not necessary for the asymptotic approach to a Friedmann-Lemaitre model. Of course the authors of Ref. \refcite{dsc} were aware of this fact as it can be read on pg. 64 where they carefully write that: if Definition 2.1 is satisfied, then the metric can be locally written in an almost-RW form.

As a final word we will use a different criteria, namely, inspired in (\ref{psi,H,assintoticos}), the asymptotic tetrad components of the Weyl and Ricci tensor
\begin{eqnarray}
&&\lim_{t\rightarrow \infty}C_{abcd}<<\epsilon \nonumber\\
&&\lim_{t\rightarrow \infty}R_{ab}\rightarrow \mbox{const.}
\label{criterio}
\end{eqnarray}
where $|\epsilon|<<1$, and the const. values do not have to vanish,  which is less restrictive than (\ref{cellis}). We emphasize that the asymptotic behavior of the tetrad components (\ref{criterio}) is a weaker criteria than the one given in Definition 2.1 of Ref. \refcite{dsc}. 

For instance, Kasner's \cite{Kasner} exact solution and the exact Bianchi $VII_A$ gravity wave of Ref. \refcite{ondaexata} obey (\ref{criterio}) with $R_{ab}\equiv 0$: on the other hand these two exact solutions do not satisfy (\ref{cellis}). So in principle our criteria would not distinguish between Kasner exact anisotropic solution, and for example, the limit  as $t\rightarrow \infty$ given in (\ref{cexassint}), with (\ref{fcexemplo}) and (\ref{cexemplo}), which is clearly a space for which isotropisation occurs in the strong sense. 

\section{Numerical Solutions}
\subsection{Asymptotically de Sitter Solution}
We choose the an initial condition near the exact de Sitter (\ref{soldeSitter}) solution with $t=0$ and $C=1$, $\alpha=1$, $\beta=-5.0$ and $\Lambda=0.06$. The only non null coefficients consistent with the 
$E_{44}\equiv 0$ constraint are  
\begin{eqnarray}
&&a_1(t)=C + 0.1\nonumber\\
&&a_2(t)=C\nonumber\\
&&a_3(t)=C\nonumber\\ 
&&a_4(t)=0.2\nonumber\\
&&\dot{a_1}(t)=C\sqrt{\Lambda/6}\nonumber\\
&&\dot{a_2}(t)=C\sqrt{\Lambda/6}\nonumber\\
&&\dot{a_3}(t)=C\sqrt{\Lambda/6}\nonumber\\
&&\ddot{a_1}(t)=C\Lambda/6\nonumber\\
&&\ddot{a_2}(t)=C\Lambda/6\nonumber\\
&&\ddot{a_3}(t)=C\Lambda/6\nonumber\\
&&\dddot{a}_1(t)= C(\Lambda/6)^{3/2}\nonumber\\
&&\dddot{a}_2(t)= C(\Lambda/6)^{3/2}\nonumber\\
&&\dddot{a}_3(t)=  8.31792283\times 10^{-4}\label{cndl}
\end{eqnarray}

\begin{figure}
 \begin{center}
 \resizebox{\imsize}{!}{\includegraphics{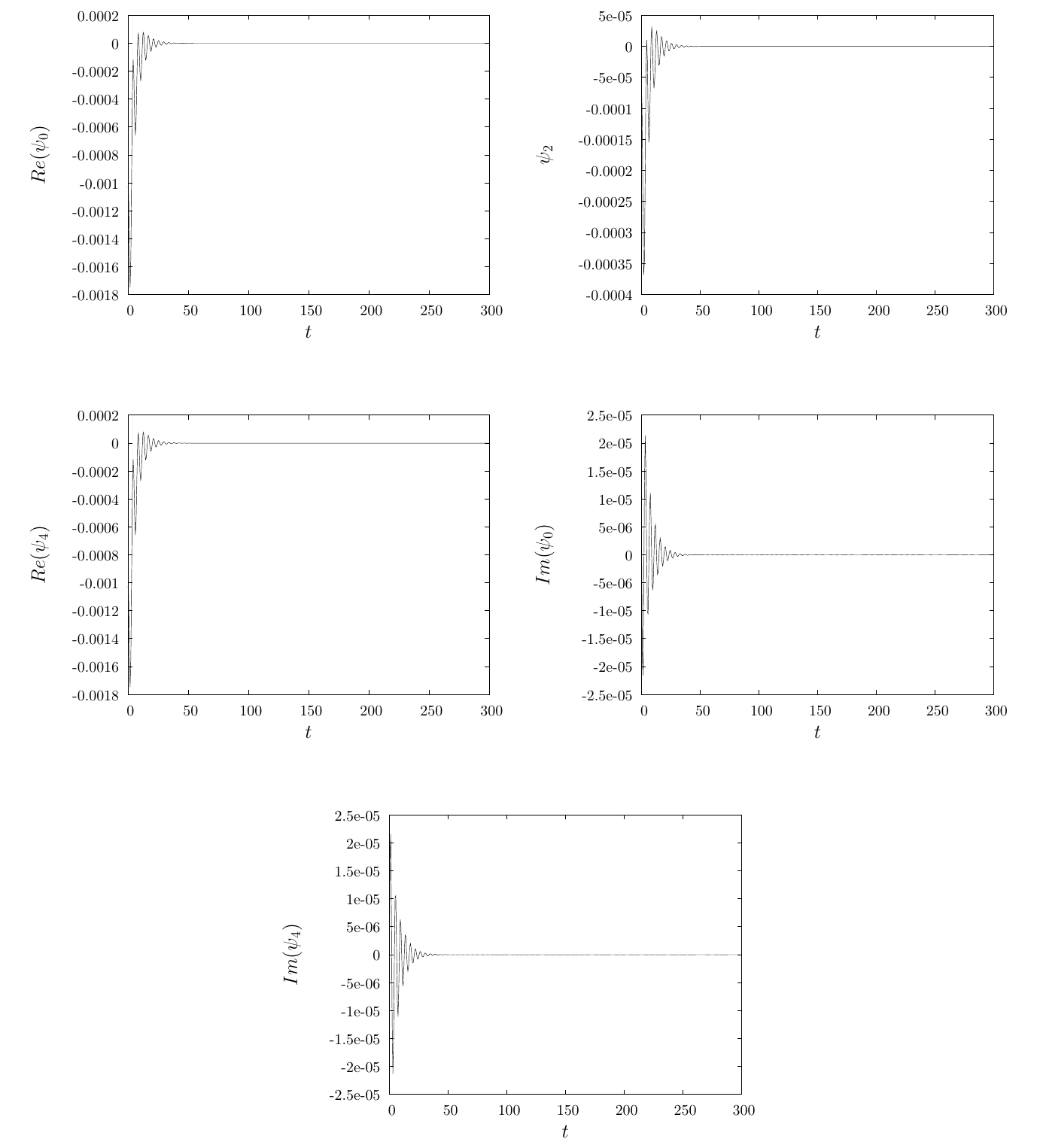}}
  \end{center}
  \caption{Numerical evolution of the real and imaginary parts of the Newman-Penrose coefficients $\psi_0$ - $\psi_4.$ For the initial condition given in the text. The numerical integration was done until the proper time $t=300$.}
  \label{figura1l}
\end{figure}

The time evolution of the non zero Newman-Penrose coefficients, $\psi_0$-$\psi_4$ are shown in Fig.  \ref{figura1l}. For the line element chosen $\psi_1\equiv \psi_3\equiv 0$. It can be seen that asymptotically the Newman-Penrose coefficients all vanish so the Weyl tensor is asymptotically zero. 

\begin{figure}
  \begin{center}
   \resizebox{\imsize}{!}{\includegraphics{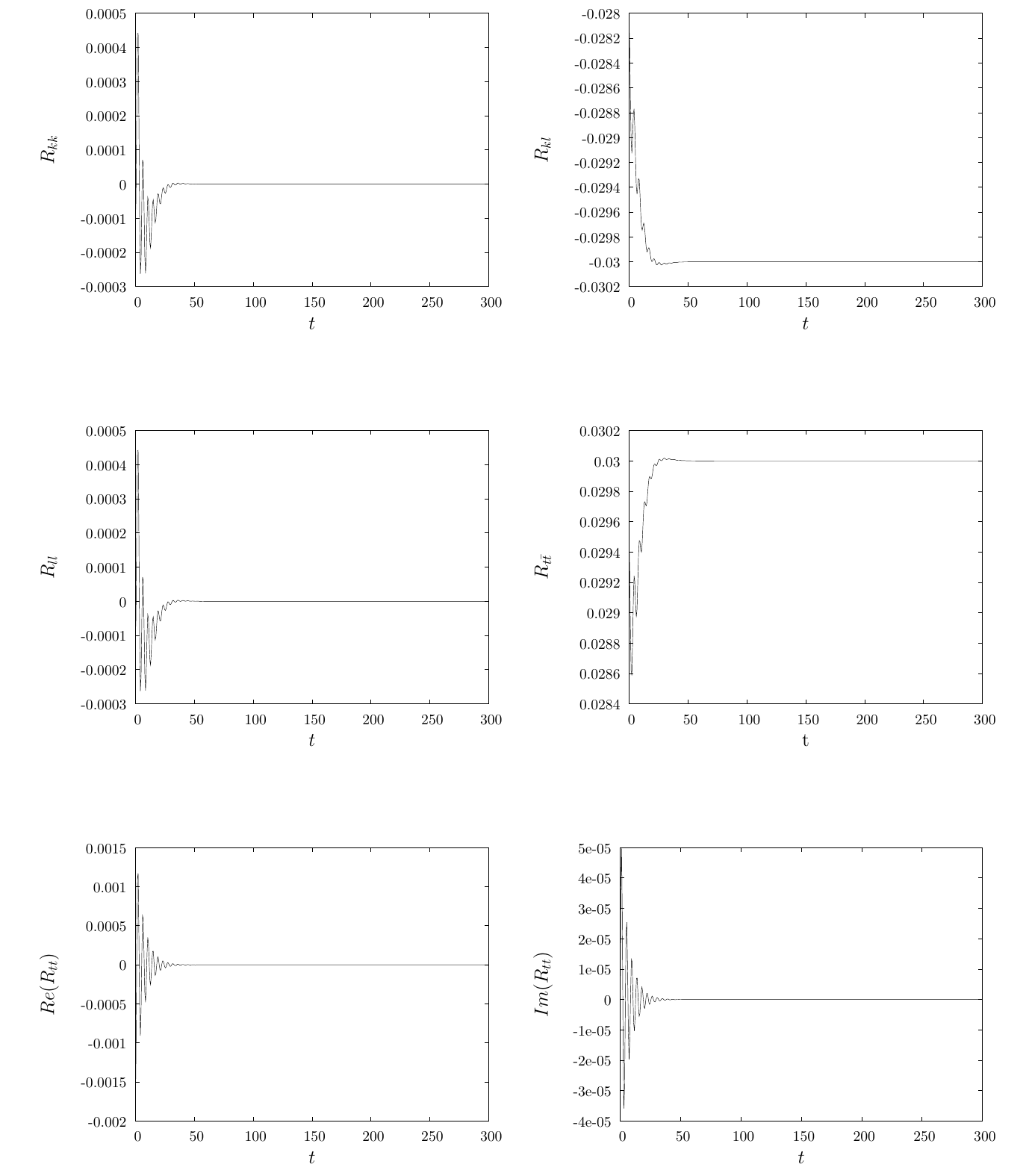}}
  \end{center}
  \caption{Numerical evolution of the non zero real and imaginary parts of the null tetrad (\ref{basenula}) components of the Ricci tensor (\ref{riccinull}). The numerical integration was done until the proper time $t=300$. It can be seen that the Ricci tensor is asymptotically proportional to the null metric (\ref{metricanula}).}
  \label{ltricci}
\end{figure}

\begin{figure}
    \begin{center}
  \resizebox{\halfsize}{!}{\includegraphics{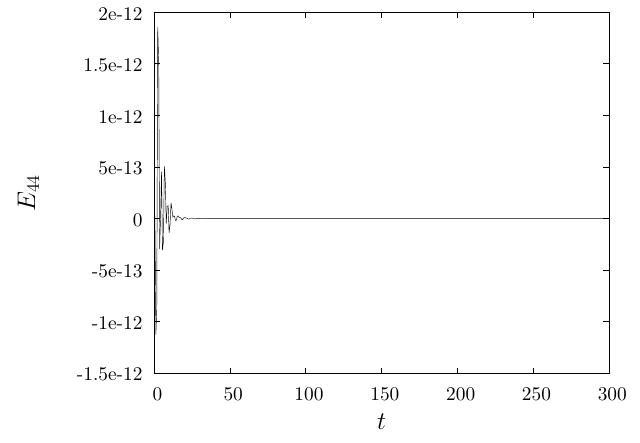}}
 \end{center}
  \caption{Numerical evolution of the constraint $E_{44}$, indicating that the numerical solution is accurate. The integration was done for the initial condition and parameters specified on the text and until the proper time $300$.  }
  \label{figura2l}
\end{figure}

The time evolution of the non zero components of the Ricci tensor according to the null base (\ref{basenula})are shown in Fig.  \ref{ltricci}. For the particular metric chosen, (\ref{eq homogeneous and anisotropic line element}) -(\ref{eq line element for spatially homogeneous four-space again}), $R_{kt}=R_{lt}=0$. 
It is also shown in Fig.  \ref{ltricci} that asymptotically, the Ricci tensor is a constant proportional to the null metric (\ref{metricanula}). The only consistent result with the field equation (\ref{eq field equation for semiclassical theory}) is 
\newpage
\[  
R_{ab}=\frac{\Lambda}{2}\left(\begin{array}{cccc}
0&-1 & 0 & 0\\
-1& 0 & 0&0\\
0 & 0&0&1\\
0&0 &1&0
\end{array}\right).
\]
\newpage
We have numerically checked up to $t=500$ in proper time that $R_{ab}=\Lambda/2 g_{ab}$ with one part in $10^{12}$. Also the constraint $E_{44}=0$, is numerically verified in Fig.  \ref{figura2l} which is a strong indication that the numerical result should be trusted. 

\subsection{Asymptotically Minkowski Solution}
In the numerical solutions the following values for the parameters $C=1.0$, $\alpha=2.0$, $\beta=-5.0$, $\Lambda=0$, where chosen, and initial conditions ($t=0$) near Minkowski solution
\begin{eqnarray}
&&a_1(t)=C+2\nonumber\\
&&a_2(t)=C\nonumber\\
&&a_3(t)=C \nonumber\\
&&a_4(t)=2.0
\label{condini}
\end{eqnarray}

The only non null initial conditions for the derivatives are given in Table \ref{tabela2}, and as before, the numerical value of $\dddot{a}_3$ is consistent with the constraint $E_{44}\equiv 0$.
 \begin{table}
 \tbl{The only non null initial conditions together with (\ref{condini}), consistent with the $E_{44}$ constraint }
{\begin{tabular}{@{}ccc@{}}\toprule
$\dot{a}_1(0)$ & $\dot{a}_3(0)$  & $\dddot{a}_3$\\ 
\colrule
$1.0$ & $0.2$  &   $-2.45214348\times 10^{-2}$,\\ 
\botrule
\end{tabular} \label{tabela2}}
\end{table}

First of all, linearization of (\ref{eq field equation for semiclassical theory}) on a slowly varying background geometry, $\dot{\delta}_{ab}<<\omega$ gives the following eigenvalues 
\begin{eqnarray}
&&g_{ab}=\eta_{ab} + \delta_{ab}e^{i\omega t}\nonumber\\
&&\omega=0,\;\;\omega=\frac{1}{\sqrt{\alpha}},\;\;\omega=\frac{1}{\sqrt{-6\beta}}, 
\label{frequencia}
\end{eqnarray}
in Riemann local coordinates where the components $|\delta_{ab}|<<1$.

\begin{figure}
  \begin{center}
   \resizebox{\imsize}{!}{\includegraphics{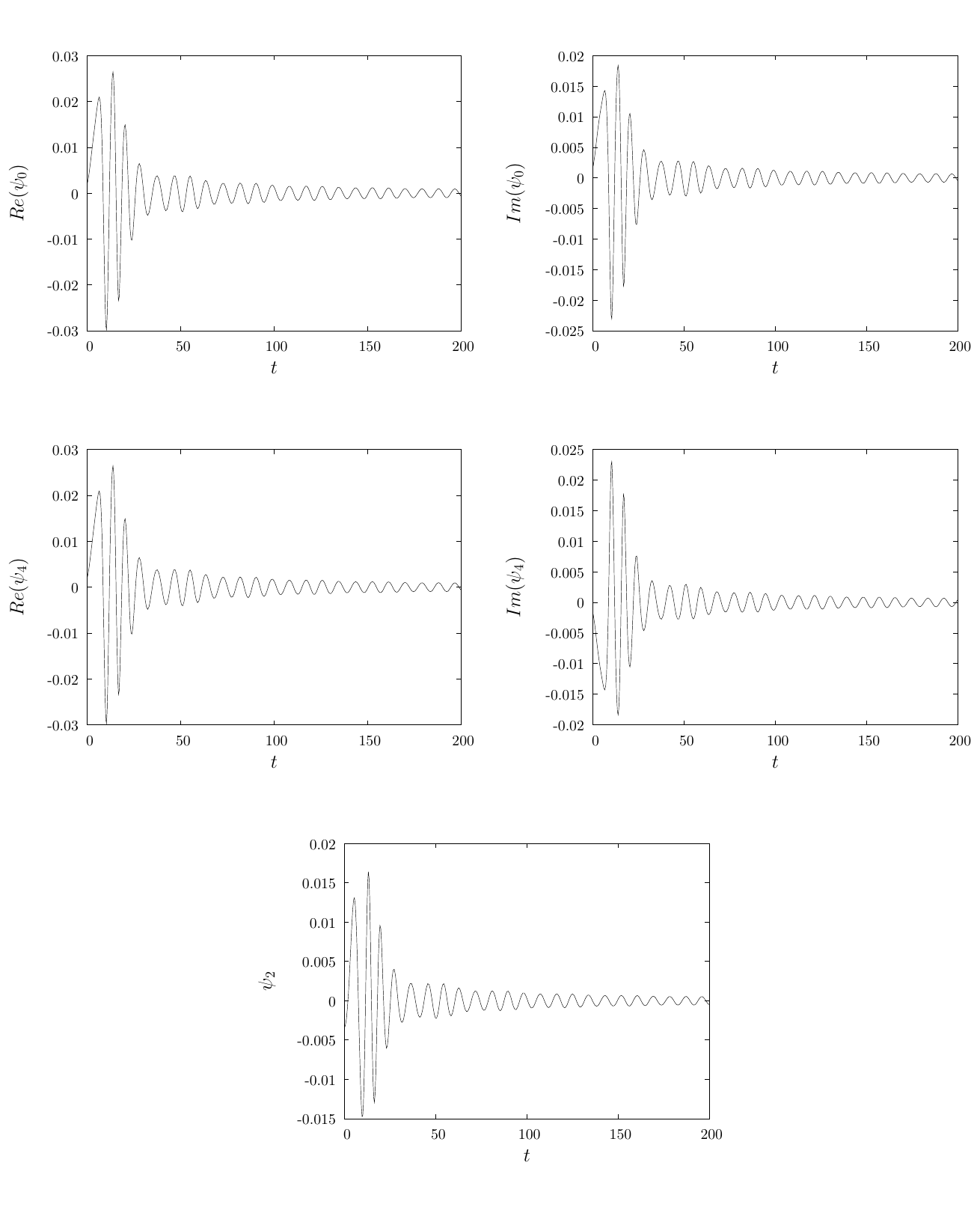}}
  \end{center}
  \caption{Numerical evolution of the real and imaginary parts of the Newman-Penrose coefficients $\psi_0$ - $\psi_4.$ For the initial condition given in the text. The numerical integration was done until the proper time $t=200$. It can be seen directly from the graphs that $\omega=2\pi/T\approx 0.706$, in good agreement with (\ref{frequencia}) $\omega=1/\sqrt{\alpha}=1/\sqrt{2}\approx0.707$}
  \label{figura1}
\end{figure}

\begin{figure}
  \begin{center}
   \resizebox{\imsize}{!}{\includegraphics{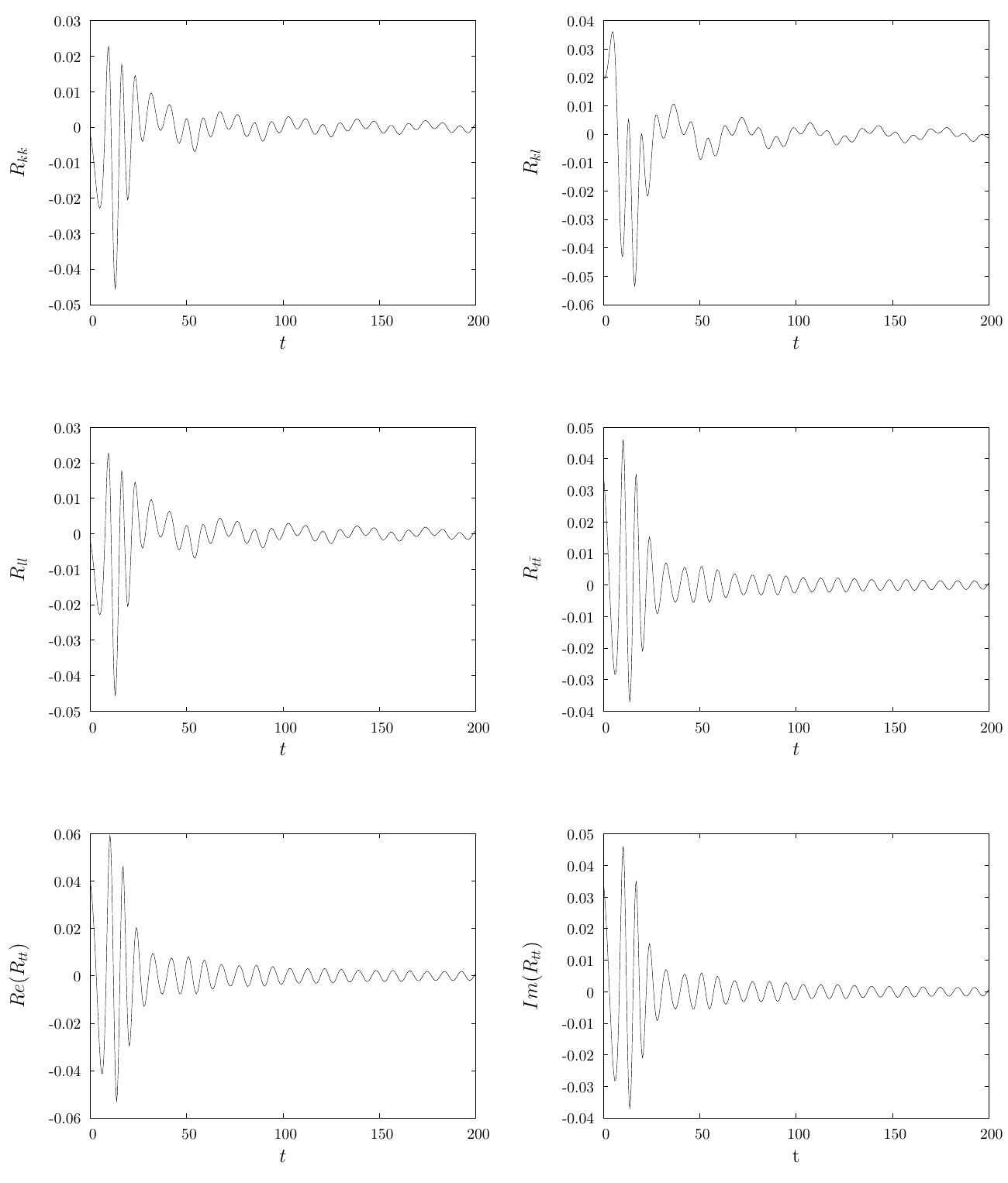}}
  \end{center}
  \caption{Numerical evolution of the non zero real and imaginary parts of the null tetrad (\ref{basenula}) components of the Ricci tensor (\ref{riccinull}). The numerical integration was done until the proper time $t=200$. Again, form the graphs the higher frequency $\omega=0.707$. We have specifically checked that the slower frequency $\omega=2\pi/T\approx 0.1820$ is also in good agreement with (\ref{frequencia}), $\omega=1/\sqrt{-6\beta}\approx 0.1825$.}
  \label{tricci}
\end{figure}

The time evolution of the non zero Newman-Penrose coefficients, $\psi_0$-$\psi_4$ are shown in Fig.  \ref{figura1}. For the line element chosen $\psi_1\equiv \psi_3\equiv 0$.
The time evolution of the non zero components of the Ricci tensor according to the null base (\ref{basenula}) are shown in Fig.  \ref{tricci}. For the particular metric chosen, (\ref{eq homogeneous and anisotropic line element}) -(\ref{eq line element for spatially homogeneous four-space again}), $R_{kt}=R_{lt}=0$. 

\begin{figure}
    \begin{center}
     \resizebox{\halfsize}{!}{\includegraphics{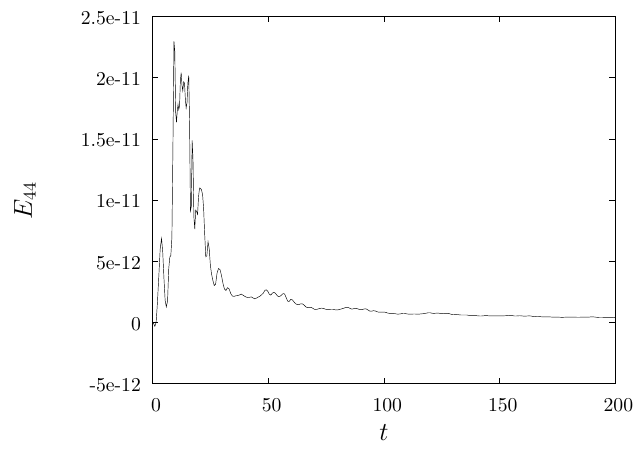}}
  \end{center}
  \caption{Numerical evolution of the constraint $E_{44}$, indicating that the numerical solution is accurate. The integration was done for the initial condition and parameters specified on the text and until the proper time $t=200$.  }
  \label{figura2}
\end{figure}

The time evolution of the constrain $E_{44}$ is shown in Fig.  \ref{figura2}.  

Since all the $\psi\rightarrow 0$, Fig.  \ref{figura1} the Weyl tensor $C^a_{bcd}\rightarrow 0$, and since the Ricci tensor $R_{ab}\rightarrow 0$ Fig.  \ref{tricci}, then the Riemann tensor $R_{abcd}\rightarrow 0$.
    
The numerical integrations shown in Fig. \ref{figura1} and Fig. \ref{tricci} were carried up to times $t=1\times 10^6$ showing that null tetrad components of the Riemann tensor $R_{abcd}\rightarrow 0$ in one part in $1\times 10^{6}$ and smaller values. It is in this sense that the solution is understood to asymptotically approach Minkowski space. 
\subsection{Singularity}
Using exactly the same parameters of the preceding section  $C=1.0$, $\alpha=2.0$, $\beta=-5.0$, $\Lambda=0$, a slightly different initial condition is chosen 
\begin{eqnarray}
&&a_1(t)=C+2.0\nonumber\\
&&a_2(t)=C\nonumber\\
&&a_3(t)=C\nonumber\\
&&a_4(t)=2.0,
\label{condsing}
\end{eqnarray}
and the only non null initial conditions for the derivatives are given in Table \ref{tabela3}. 
 \begin{table}
 \tbl{The only non null initial conditions together with (\ref{condsing}), consistent with the $E_{44}$  constraint}
{\begin{tabular}{@{}cccc@{}}\toprule
$\dot{a}_3(0)$ &  $\dddot{a}_3$\\ 
\colrule
 $2.0$  &  $-7.111111$,\\ 
 \botrule
\end{tabular} \label{tabela3}}
\end{table}

\begin{figure}
  \begin{center}
   \resizebox{\halfsize}{!}{\includegraphics{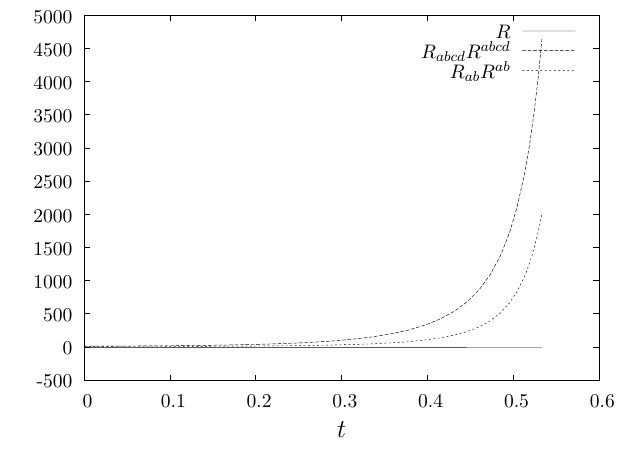}}
    \end{center}
    \caption{The increase of the scalar curvature invariants $R_{ab}R^{ab}$, $R_{abcd}R^{abcd}$ characterizing a singularity.
    \label{sing}}
    \end{figure}

\begin{figure}
  \begin{center}
   \resizebox{\halfsize}{!}{\includegraphics{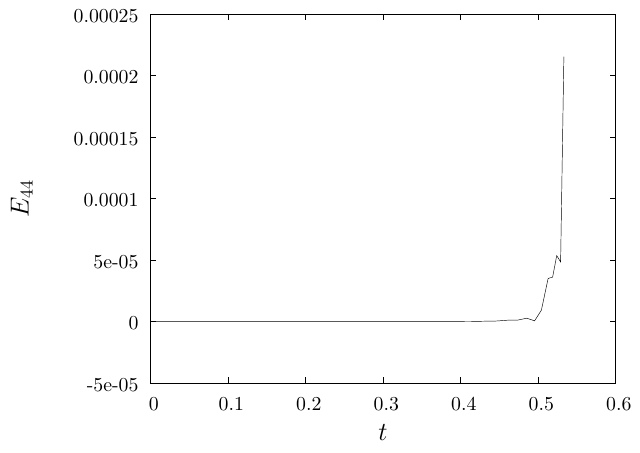}}
    \end{center}
    \caption{The constraint $E_{44}$. Note that as the singularity is approached, the numerical errors increase, which is expectable. 
    \label{controle}}
    \end{figure}

This initial condition evolves very fast to a singularity characterized by the increase of the curvature scalars $R_{ab}R^{ab}$ and $R_{abcd}R^{abcd}$ shown in Fig.  \ref{sing}.
The constraint which should be zero $E_{44}$ shown in Fig.  \ref{controle}, prove that the numerical result is accurate. Note that as the singularity is approached, the numerical errors increase, which is expectable. Also, the numerical integration can be carried further and further, and the constraints are not really satisfied at all, and is not shown since we believe the result should not be trusted. 
    \newpage
\section{Conclusions}
In this present work, the numerical solutions of Bianchi $I$ type are analyzed in the effective gravity context. This is the anisotropic generalization 
of the Euclidean $E^3$ Friedmann ``flat" space. The quadratic gravity is the result of vacuum polarization counter terms which must be introduced into Einstein's theory of gravitation. It should be the most natural theory just after the Planck era. It must be emphasized that although numeric the solutions are exact in the sense that they depend only on the precision of the machine. 

We have previously,  numerically investigated Bianchi $I$ type solutions \cite{sandro}. We found that there is a basin of attraction to Minkowski space for $\Lambda=0$ and a basin of attraction to de Sitter space for $\Lambda>0$, see \cite{daniel-sandro}. In this sense, Minkowski and de Sitter solutions are structurally stable according to the effective gravity, for the particular Bianchi $I$ models we analyzed. 

The numerical solutions are given in section 3. In section 3.1 the initial condition is chosen near the de Sitter exact solution. In section 3.2. and 3.3 the the solution is chosen near Minkowski exact solution. 

The solutions are understood asymptotically in the following sense. The Riemann tensor depends on the Weyl tensor, Ricci tensor and Riemann scalar. The numerical solutions are characterized asymptotically according to the null tetrad components of the above tensors. 

In this sense, for the first solution under consideration in section 3.1, the Weyl tensor is zero and the Ricci tensor is proportional to the metric $R_{ab}=\Lambda/2g_{ab}$ asymptotically and we identify this solution with de Sitter space. In section 3.2 the solution asymptotes Minkowski space in sense that the Weyl tenosr and Ricci tensor vanish asymptotically. The convergence to Minkowski space is not so fast and we shall not address this question in this present work. In section 3.3 a solution that converges to a singularity is presented. 

In all the cases we considered, the numerical behavior described in the text was checked for much larger times that the ones plotted, and we believe that asymptotical interpretation is the correct one. 

So this work is in accordance with previous results obtained by on us for the more particular case of the  diagonal Bianchi $I$ case \cite{daniel-sandro,sandro}. As in the diagonal metric case, note the presence of the singularity, depending on the initial condition: this theory certainly can not be understood as a complete one because an initially reasonable condition evolves to a singular universe. On the other hand, depending on the initial condition isotropisation occurs in a weak sense, see discussion on pg. 3 of this work.

Since the initial conditions are near de Sitter and Minkowski solutions, we can speculate that also for Bianchi $I$ solutions,  de Sitter and Minkowski space should be structurally stable according to the effective gravity for the non diagonal metric also,  in the sense that there should be basins of attraction to these solutions. We intend to address this stability issue in a future work. 

\section*{Acknowledgments}
J. A. de Deus wishes to thank the Brazilian agency CNPq for
financial support. D. M. wishes to thank the Brazilian projects: {\it
Nova F\'\i sica no Espa\c co} and INCT-A.

\label{lastpage}
\end{document}